# Advances in Nb₃Sn superconducting radiofrequency cavities towards first practical accelerator applications

S. Posen[1a)], J. Lee[1,2], D.N. Seidman[2,3] A. Romanenko[1], B. Tennis[1], O. S. Melnychuk[1], D. A. Sergatskov[1]

[1]Fermi National Accelerator Laboratory, Batavia, IL, 60510, USA
[2]Department of Materials Science and Engineering Northwestern University, Evanston, IL, 60208, USA
[3]Northwestern University Center for Atom-Probe Tomography (NUCAPT), Evanston, IL, 60208, USA

Nb₃Sn is a promising next-generation material for superconducting radiofrequency cavities, with significant potential for both large scale and compact accelerator applications. However, so far, Nb₃Sn cavities have been limited to cw accelerating fields <18 MV/m. In this paper, new results are presented with significantly higher fields, as high as 24 MV/m in single cell cavities. Results are also presented from the first ever Nb₃Sn-coated 1.3 GHz 9-cell cavity, a full-scale demonstration on the cavity type used in production for the European XFEL and LCLS-II. Results are presented together with heat dissipation curves to emphasize the potential for industrial accelerator applications using cryocooler-based cooling systems. The cavities studied have an atypical shiny visual appearance, and microscopy studies of witness samples reveal significantly reduced surface roughness and smaller film thickness compared to typical Nb₃Sn films for superconducting cavities. Possible mechanisms for increased maximum field are discussed as well as implications for physics of RF superconductivity in the low coherence length regime. Outlook for continued development is presented.

## I. Introduction

Superconducting radiofrequency (SRF) cavities are widely used for particle beam acceleration because of their ability to generate large amplitude accelerating electric fields ($E_{acc}$) with small heat dissipation ($P_d$). Niobium has been the material of choice for SRF cavities for decades, having the highest critical temperature ($T_c$) of the elements, ~9.2 K, allowing it to have a high quality factor $Q_0$ (small $P_d$) at temperatures accessible with liquid helium, with typical operation ~2 K. Nb₃Sn is a new SRF material that is being explored, with significantly higher $T_c$ (~18 K) than Nb (~9 K), which has allowed it to achieve higher $Q_0$ than Nb in a wide temperature range [1]. However, Nb₃Sn cavities have been limited to cw (continuous wave, i.e. not pulsed) accelerating gradients significantly smaller than Nb, with the best cavities being limited by quench (loss of superconductivity) at 18 MV/m, compared to the best Nb cavities reaching up to ~50 MV/m [2]. Furthermore, larger Nb₃Sn cavities—cavities with more than 2 cells or with frequency below ~1 GHz—have in past research programs had limited development and relatively poor performance in the small number of tests performed on them, generally limited to ~5 MV/m [3], [4].

Development of Nb₃Sn to overcome these limitations is motivated partially by theoretical predictions of higher maximum fields if the films are sufficiently optimized [5] and partially by the ability to operate at higher temperatures where cryogenic efficiency is far higher[1] [6]. In addition to the potential for large scale applications, researchers are developing cryocooler-based cooling schemes [7]–[9] to enable compact accelerator applications based on Nb₃Sn cavities operating at ~4 K for small- and medium-scale applications such as wastewater treatment and medical isotope production.

Achieving a uniform stoichiometry over the RF surface of an SRF cavity is significantly more complex with Nb₃Sn than with Nb. The superconducting properties of Nb-Sn compounds vary with composition, and there is a narrow composition range around 25% Sn with the highest $T_c$ [10]. There are a number of methods to create Nb₃Sn, but so far, the method that has been most successful in terms of cavity performance is vapor diffusion [4], [11], [12], which takes advantage of the phase diagram at high temperatures to "phase-lock" to the desired composition (as described in Ref. [13]). This technique is now used by several labs around the world for coating cavities with Nb$_3$Sn [1], [14], [15]. Researchers continue to optimize the process through a combination of coating parameter space exploration, cryogenic RF measurement, microstructural analysis, and materials science. There has been significant recent progress in understanding detrimental microstructural imperfections that can occur in vapor diffusion Nb$_3$Sn coatings, including abnormally large thin grains (so-called "patchy" regions) [16]–[18] and grain boundaries (GBs) with segregated tin [19]. It has been demonstrated that these imperfections in Nb$_3$Sn can be minimized by carefully selecting growth parameters.

---

[1] e.g. a large cryogenic plant has a coefficient of performance approximately 3-4 times better at 4.4 K vs 2 K.

In this paper, we report on experiments exploring modifications to typical coating procedures to make films that have relatively low surface roughness and relatively small film thickness. Reducing surface roughness may augment the energy barrier to flux penetration and therefore increase the quench field [20]. Thinner films may have higher quench field due to the poor thermal conductivity of Nb$_3$Sn, ~$10^3$ lower than that of Nb [21], for example by improving stability in the case of localized defects. We show measurements of several types of SRF cavities with smooth, thin films, showing significantly improved performance compared to the previous state-of-the-art.

## II. Nb$_3$Sn Coating Process Development

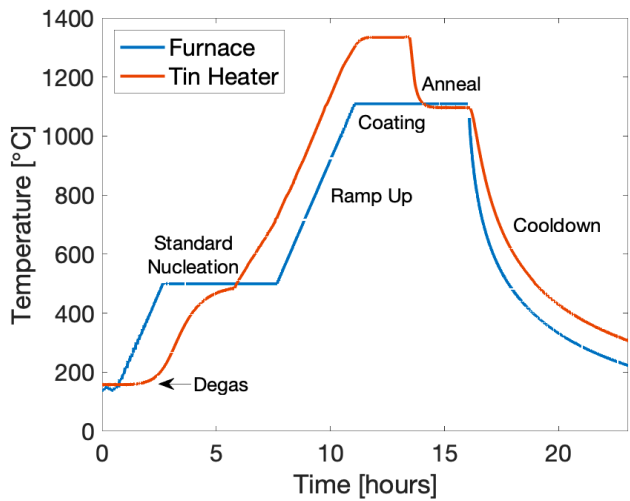

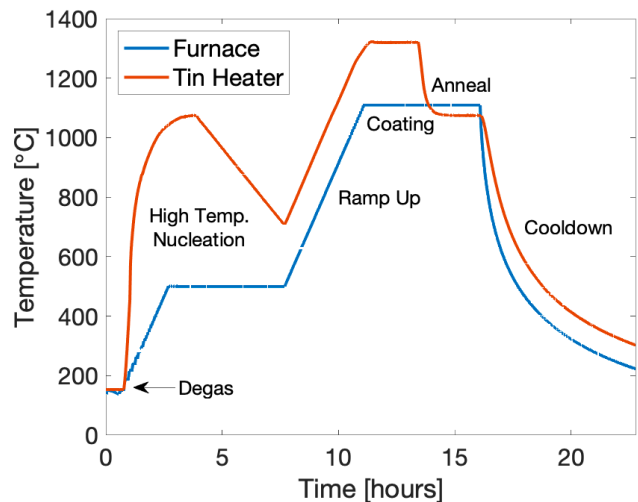

**Fig. 1**: Vapor diffusion coating process at Fermilab. Top: with standard nucleation step at ~500°C. Bottom: with modified high-temperature nucleation step.

A typical vapor diffusion Nb$_3$Sn coating procedure is shown in the top of **Fig. 1**. The process involves 6 stages (described in more detail in [1]):
1) Degas at ~140°C – allows ramp up to proceed without large pressure spikes. Typical duration ~24 hours (not shown in **Fig. 1**).
2) Nucleation at ~500°C – SnCl$_2$ has a high vapor pressure in this temperature range and creates Sn sites on the surface
3) Ramp up – Furnace is ramped up to coating temperature ~1100°C. Sn heater is used to maintain the Sn source ~100-200°C hotter than the furnace to maintain high vapor pressure.
4) Coating at ~1100°C – Sn evaporates and reacts with the Nb of the substrate to form Nb$_3$Sn
5) Anneal at ~1100°C – the Sn heater is turned off but the furnace is maintained at temperature to give a chance for excess Sn on the surface to evaporate or diffuse
6) Cooldown

Steps 2 and 3 to maintain high vapor pressure were recommended by Hillenbrand et al. [22] and a proposed explanation for the benefit provided by the high vapor pressure is reducing the formation of anomalously large thin grains [18].

Procedures for producing low surface roughness, thin Nb$_3$Sn coatings are not yet well established and repeatable. Smooth, thin films have been produced several times at Fermilab, and in this paper, we will present results from them, but additional studies are needed to fully understand which coating parameters are key determinants for these properties. However, we can identify several common features among coating procedures that have produced smooth, thin films:

- The niobium cavity substrates were given electropolish treatment (as opposed to buffered chemical polish) to achieve a smooth surface prior to coating
- The niobium substrates were anodized to 30 V in ammonia prior to coating (recommended previously, e.g. [22] and [23])
- To encourage high vapor pressure, the Sn heater was driven with maximum power available (measured ~1300°C in thermocouples in heater coil, expect somewhere between 1200°C-1250°C in Sn crucible based on previous calibration)
- To encourage high vapor pressure, a relatively large crucible diameter was used (~15 mm or larger)
- To prevent condensation of Sn droplets on the surface due to a high vapor pressure in a closed volume, one or more ports of the cavity were kept open to the chamber (similar to the Cornell setup [1])
- The nucleation step was substantially modified, to have a rapid ramp to high temperatures ~1000°C – this will be discussed in detail below
- A nitrogen infusion step was added at the end of the coating process – this is also discussed in detail below

The high temperature nucleation step is illustrated in the bottom of **Fig. 1**. Immediately following the degas, while the furnace temperature begins a slow ramp to 500°C, the Sn source heater is driven with high power to raise its temperature very quickly, achieving eventually a maximum temperature of ~1000°C. The heater raises the temperature of two crucibles, one that contains Sn and one that contains $SnCl_2$. The vapor pressure of the Sn is still expected to be relatively low at these temperatures, but the $SnCl_2$ is expected to have a very high vapor pressure [24]. It is expected that the full quantity of $SnCl_2$ will evaporate very quickly, much faster than it would during the other process shown in the top of **Fig. 1**, especially considering the observed thermal lag of the unpowered Sn heater relative to the furnace. The goal of the high temperature nucleation step is to create a very high vapor pressure to a) reduce the mean free path of the vapor and thereby improve homogeneity over the surface, and b) rapidly transport Sn to the surface early in the process (see [18], [25]).

The nitrogen infusion step was added in an attempt to raise maximum gradients as has been demonstrated in niobium cavities [26]. Once the furnace temperature has dropped to 120°C, that temperature is maintained for 48 hours, and nitrogen gas is bled into the furnace, maintaining ~25 mtorr of pressure (see **Fig. 2**). In niobium cavities, this process creates nitrogen interstitials in the near surface, and it is hypothesized that the resulting "dirty" layer in the superconductor enhances the energy barrier preventing flux entry into the superconductor.

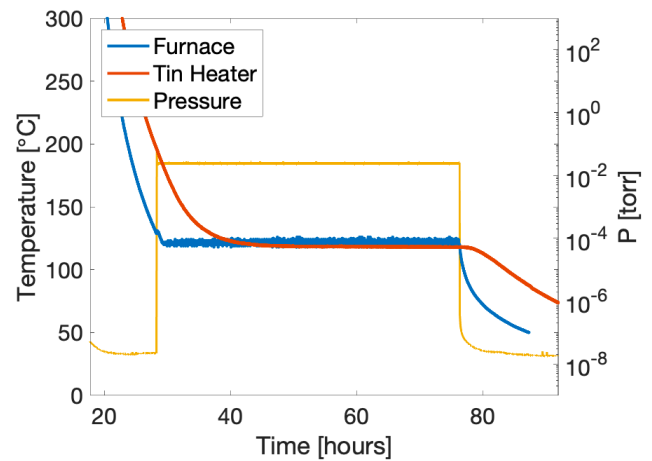

**Fig. 2**: Example of a N-infusion step after $Nb_3Sn$ coating

A number of coatings were performed, varying parameters described above as well as others such as tin quantities. In all cases, cavities were evaluated at least visually, and sometimes also with cryogenic testing. Visual evaluation would often give useful indications of coating quality. Examples include patterns of dark spots (interpreted as tin condensation) [18], local shiny spots (interpreted as anomalously large thin grains), and a shiny, almost glossy appearance over the whole surface (interpreted as a low surface roughness coating).

Recently, some repeatability has been established in generating shiny, glossy films by reducing the quantity of tin in the coatings substantially (by ~50%). Providing less tin from the source would be consistent with the films being thinner. However, this reduction of tin quantity was not implemented for the first shiny coatings produced at Fermilab, so it is not clear if it is a requirement. The recent repeatable shiny films have not been evaluated with cryogenic testing due to limitations caused by the COVID-19 pandemic. This will be a focus of future paper; in this work we will focus on the results already obtained.

### III. Microscopy of Nb$_3$Sn coatings

Nb$_3$Sn samples were analyzed using optical, laser confocal, scanning electron (SEM), and transmission electron microscopy (TEM). Surface roughness was characterized using Bruker ContourGT Optical profiler. SEM imaging of Nb$_3$Sn samples was performed using FEI Quanta SEM with 30kV electron beam and TEM analyses was performed using Hitachi HD 2300 electron microscopes with 200 kV electron beam.

Typically, Nb$_3$Sn vapor diffusion coatings have a matte, nonreflective gray visual appearance. However, the cavities studied in this paper had a shiny, reflective appearance immediately noticeable upon removal from the furnace. Witness samples coated with the cavities also show a shiny appearance, and surface roughness measurements via laser confocal microscope show a significantly smoother surface than typical coatings, consistent with the shiny appearance. This is illustrated in **Fig. 3**.

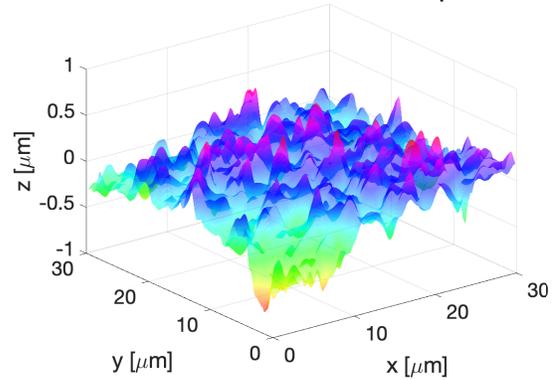

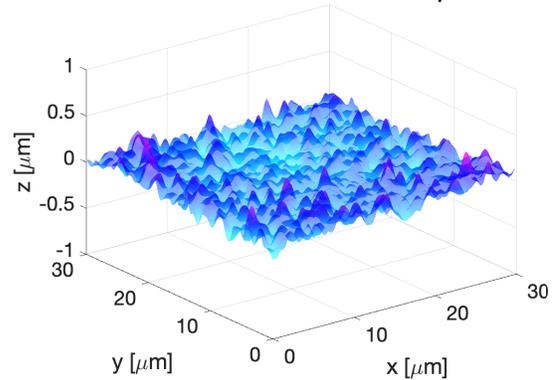

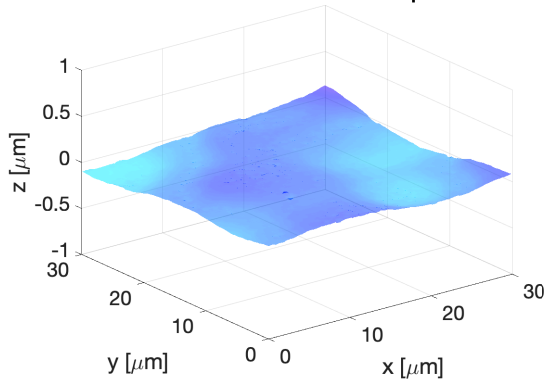

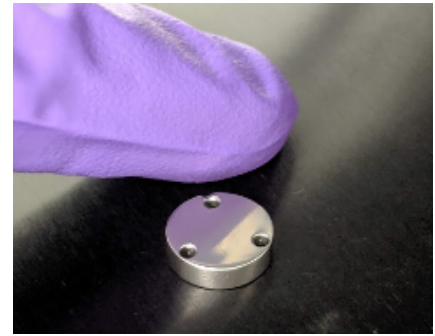

**Fig. 3**: Comparison of surface roughness (represented by arithmetic mean and root mean squared deviation R$_a$ and R$_q$) of 3 samples: electropolished niobium sample before coating (top), a coating typical of previous matte Nb$_3$Sn (second from top), and a new shiny Fermilab Nb$_3$Sn-coated sample (second from bottom). A photo shows the shiny Nb$_3$Sn sample surface reflecting a gloved finger (bottom).

**Fig. 4** compares the microstructure of a matte Nb$_3$Sn coating and that of a shiny Nb$_3$Sn coating, both plane-view and in cross-section. It shows that the shiny Nb$_3$Sn coating is thinner (~1 µm) than the matte Nb$_3$Sn coating (~2 µm). Also, its grain diameter is smaller (0.93 ±0.30 µm) than the matte Nb$_3$Sn coating (1.37 ±0.57 µm).

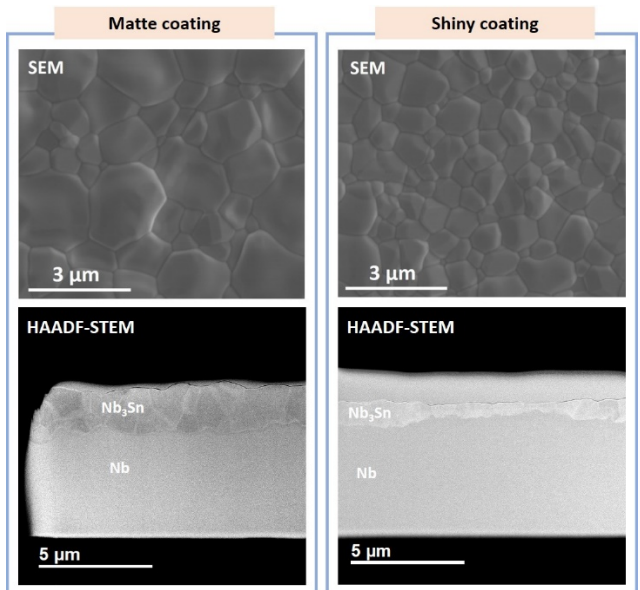
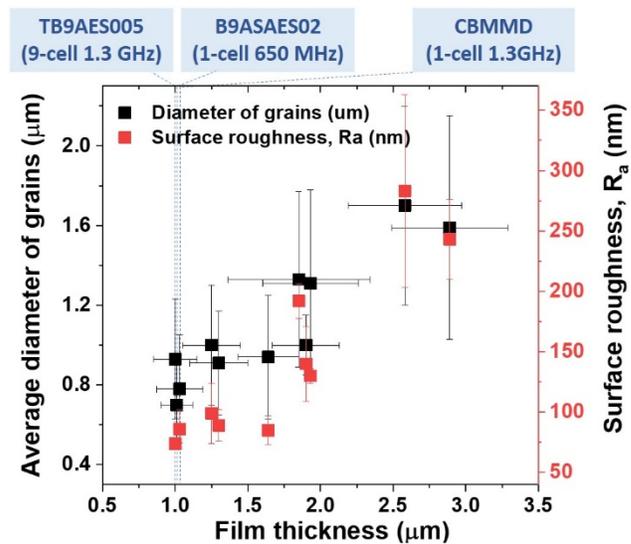

**Fig. 4**: Plane-view SEM and cross-sectional TEM image of Nb₃Sn coatings on Nb. The shiny Nb₃Sn coating is thinner (~1 μm) and has smaller grain diameter (<~1 μm) than the matte Nb₃Sn coating.

Surface roughness values of various Nb₃Sn coatings prepared by standard and new coating procedures are measured by optical microscopy and summarized in **Fig. 5**. It demonstrates that the surface roughness of the Nb₃Sn coatings is strongly correlated to the film thickness and average diameter of Nb₃Sn grains: thinner coatings have smaller grain size and smaller surface roughness. Typical matte Nb₃Sn coatings are ~2 μm thick with ~1.5 μm grain diameter, and more than ~150 nm of surface roughness ($R_a$) values, measured in a 47 μm by 62 μm area. The shiny Nb₃Sn coatings are thinner, ~1 μm, have smaller grain diameter, less than ~1 μm, and have surface roughness as low as $R_a$~80 nm. A previous study performed by adjusting coating temperature and duration showed a similar general relationship between film thickness, grain size, and surface roughness [27]. The effect of these factors on Nb₃Sn cavity performance is presented and discussed in sections IV and V.

**Fig. 5**: Correlation among film thickness, average diameter of Nb₃Sn grains, and surface roughness of Nb₃Sn coatings on Nb: it clearly shows that surface roughness is roughly proportional to the film thickness and average grain diameter of Nb₃Sn coatings.

HAADF-STEM images in **Fig. 6** display the height differences among grains as well as grooves at grain boundaries in a sample of matte Nb₃Sn. Height differences ~180 nm between Nb₃Sn grains are observed as well as grooves at GBs ~40 nm deep with ~190 nm width. Comparative analyses between the surface profiles measured by optical microscope and HAADF-STEM images of the surface morphology of Nb₃Sn coatings indicates that the height differences among Nb₃Sn grains (~200 nm) and grooves (~40 nm) are probably the primary causes of the surface roughness of Nb₃Sn coatings.

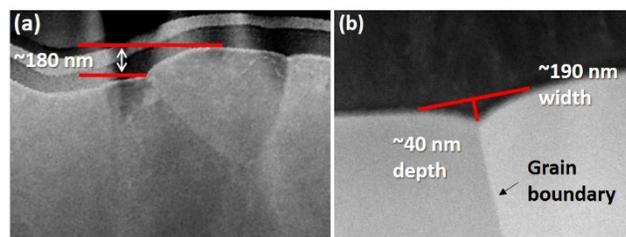

**Fig. 6**: HAADF-STEM images illustrate the causes of surface roughness in a matte Nb₃Sn coating: (a) height differences across grains, ~180 nm, and; (b) groves at grain boundaries (GBs) with ~190 nm width and ~40 nm depth.

We also note that the shiny coating has a narrower grain size distribution. **Fig. 7** displays histograms of grain diameters in two Nb₃Sn coatings: (i) matte and; (ii) shiny. The matte Nb₃Sn coating shows a wide distribution of grain diameter from 300 nm to 3 μm. In contrast, the shiny Nb$_3$Sn coatings show a narrow distribution of grain diameter around ~700 nm. The size distributions of Nb$_3$Sn grains are normalized and fit with log-normal distributions. The shiny Nb$_3$Sn coating displays significantly smaller standard deviation ($\sigma$=0.25) compared to the matte coating ($\sigma$=0.45).

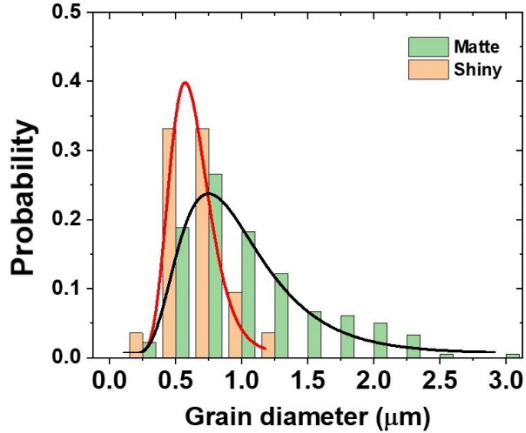

**Fig. 7**: Histogram of grain size distribution in matte and shiny Nb$_3$Sn coatings: it reveals wider distribution of Nb$_3$Sn grains in the matte ($\sigma$=0.25) compared to the shiny coating ($\sigma$=0.45), where $\sigma$ is standard deviation assuming they follow log-normal distribution.

Compositional analysis was performed on a witness sample coated with single cell 1.3 GHz cavity CBMM-D using SIMS (secondary ion mass spectrometry). The nitrogen-treated sample did *not* show a significant excess of nitrogen in the near surface compared to witness samples that did not receive a nitrogen infusion step. **Fig. 8** contrasts this with a N-infused Nb sample, that shows an excess of N relative to an EP'd sample, several nm into the near surface, below the oxide. It is therefore unclear if the nitrogen infusion step had any effect, but after observing strong RF performance in CBMM-D, it has typically been included in an effort to maintain consistency while trying to reproduce its coating procedure. Additionally, several coatings were attempted with the N-infusion step that resulted in Nb$_3$Sn layers that were not smooth, shiny, or thinner than ~2 µm, suggesting that this step was not the key factor for resulting in these qualities.

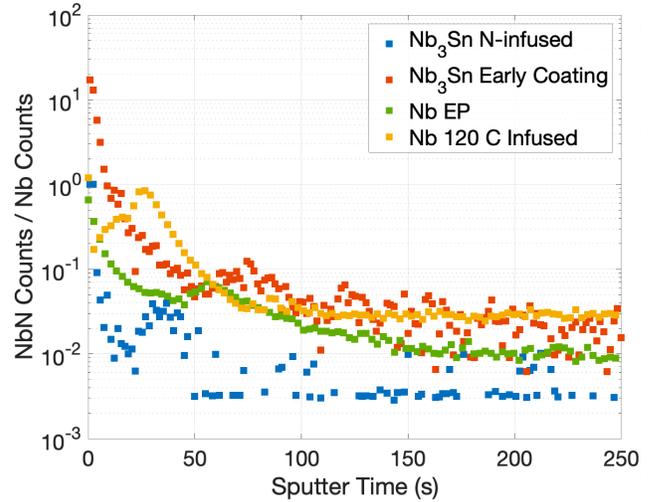

**Fig. 8**: Depth profile of nitrogen in various samples, measured by SIMS. Signals are normalized to Nb counts. Sputtering rate is approximately 0.1 nm/s.

Next, we present studies of the effect of the high-temperature nucleation step on the microstructure of Nb$_3$Sn coatings. This step is believed to increase the density of Sn nucleation sites on the surface, especially in regions that do not have line of sight to the coating source. This in turn is expected to reduce the formation of "patchy" regions on the surface [18]. The high density of nucleation sites may also contribute to the smaller grain size in the final film. The effect on nucleation was studied as a function of location in the cavity with respect to the Sn source using a sample-host cavity. **Figs. 9(a,b)** show the sample-host cavity with Nb coupon substrates placed at various locations (#1-12). A molecular flow simulation of Sn vapor using COMSOL software indicates that there is a variation of Sn number density, which represent Sn vapor pressure, inside the cavity depending on the locations with respect to the Sn source in the cavity from #1 to #12, **Figs. 9(c)** [28]. The inner surface of the cavity with line of sight to the Sn source is expected to have relatively high-Sn vapor pressure (location #5, 6 in **Fig. 9**) and the inner surface of the Nb cavity around the corner from the Sn source is expected to have lower vapor pressure (location #1, 2, 7, 8 in **Fig. 9**). These results should be considered for illustrative, qualitative purposes only as the simulation was not adjusted to match with the specific coating parameters (and moreover viscous flow may be more appropriate than molecular flow for a high temperature nucleation).

Variations were observed in the surface coverage of islands on the Nb substrate surface in the samples prepared by the standard 500°C nucleation step and the high temperature nucleation step. **Fig. 10** compares the surface islands formation on Nb surface at locations #1 and #6. The coverage appears denser and somewhat more uniform in the high temperature nucleation, with larger islands.

A number of images like those in **Fig. 10** were analyzed using imageJ software [29] to estimate the areal fraction of the surface coverage of islands on the Nb substrate. At least 3 images were analyzed in each sample from both coatings, and the mean values of areal coverage are plotted in the top of **Fig. 11**. We note that for the standard nucleation, there is a somewhat higher areal fraction in the samples from regions with high Sn number density in **Fig. 9c)**, a trend that is not observed in the high temperature nucleation samples. This increased uniformity of islands on the samples with the high temperature nucleation step is consistent with an increased Sn vapor pressure in the cavity. The average areal coverage of surface islands on Nb substrate is 9.0±0.5 % in the high temperature nucleation procedure, which is ~4 times larger than the areal fraction of islands prepared by the standard nucleation, 2.5±0.6 %. The smaller areal fraction of islands for the standard nucleation samples may be due to lower vapor pressure leading to more line-of-sight deposition on the beamtubes of the cavity, and therefore less vapor being available to reach the samples.

For sample 6 of each set of samples, 13-14 images were analyzed, to give an idea of uncertainty in the measurement, plotted in the bottom of **Fig. 11**. There are variations of areal fraction of islands from one Nb grain to another, especially for those that received the high temperature nucleation, and it causes a variation of areal fraction of surface islands even at the same location in the cavity. It may indicate that the crystal orientation of the Nb surface affects the local areal coverage of islands during nucleation, which would be consistent the study by Pudasaini et al. in Ref. [30].

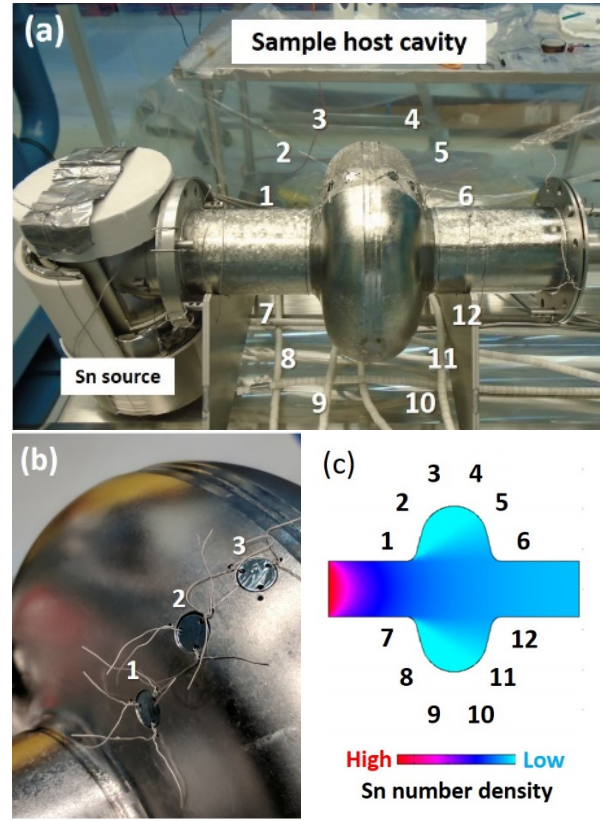

**Fig. 9**: (a) Image of sample-host cavity with Sn source for $Nb_3Sn$ coating. (b) Nb substates placed at the location #1, 2, 3 in the sample-host cavity. (c) Schematic of Sn number density in the cavity during the coating process, prepared by molecular flow simulation of Sn vapor in the cavity during the coating process using COMSOL software (reprint from ref. [28]).

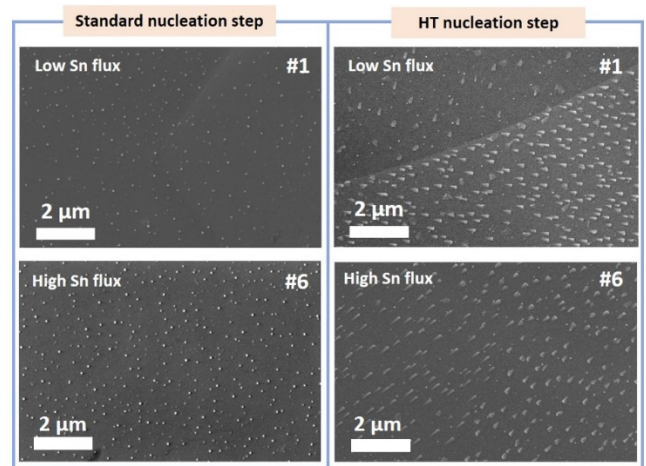

**Fig. 10**: SEM images of precipitates on Nb substrates with different Sn flux, after two different nucleation steps of $Nb_3Sn$ coatings: (i) standard nucleation step; and (ii) high-temperature nucleation step.

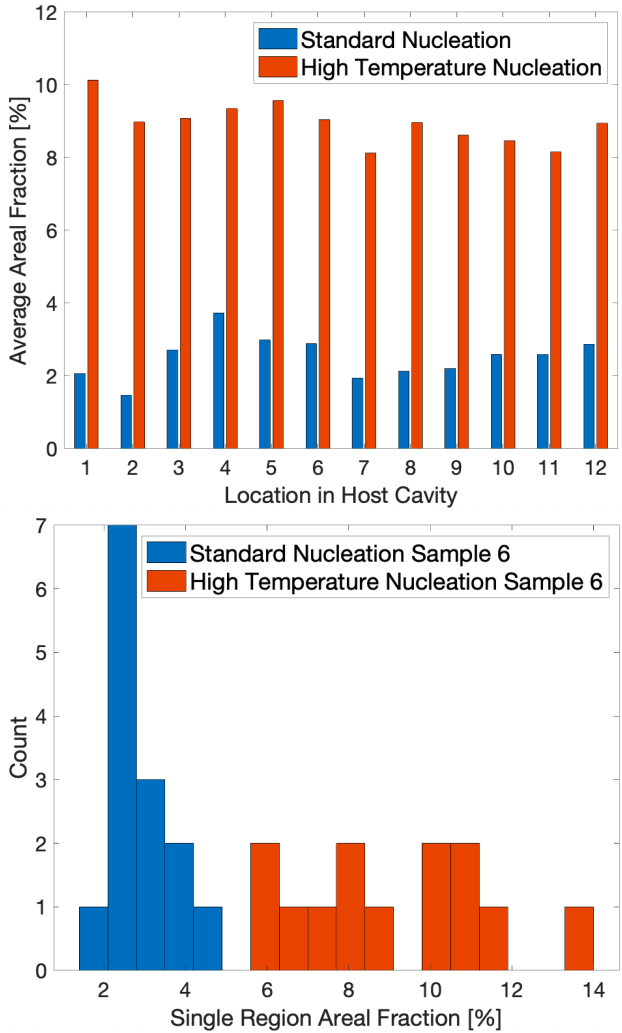

**Fig. 11**: Coverage of islands on Nb samples at different locations in the sample-host cavity. The high temperature nucleation step results in a larger and somewhat more uniform areal fraction of precipitates. The top plot shows the average of several images taken over a given sample. The bottom plot shows histograms of several images measured on sample 6 from each nucleation procedure.

## IV. Performance of Nb₃Sn SRF cavities

The performance of the cavities is evaluated in a vertical test cryostat using typical cavity RF measurement procedures (e.g. see [31]). Cavities are cooled slowly and uniformly through the critical temperature ~18 K to prevent degradation due to trapped thermocurrents [4]. Cancellation coils are used to minimize the ambient field around the cavities to ~1 mG or smaller. Tests are usually performed with slightly pressurized liquid helium at 4.4 K and with subatmospheric superfluid ~2 K. Measurements are also sometimes performed in cold helium gas to evaluate performance at higher temperatures for cryocooler-based applications.

Several cavities with the shiny film appearance were tested in liquid helium in Fermilab's vertical test stand (VTS): a single cell 1.3 GHz cavity CBMM-D, a single cell 650 MHz cavity B9AS-AES-002, and two 9-cell 1.3 GHz cavities, TB9ACC014 and TB9AES005. The performance of CBMM-D is shown in the top of **Fig. 12**. It reached 22.5 MV/m in the first VTS test, a new cw accelerating gradient record for Nb₃Sn cavities, which previously have been limited to ~18 MV/m.

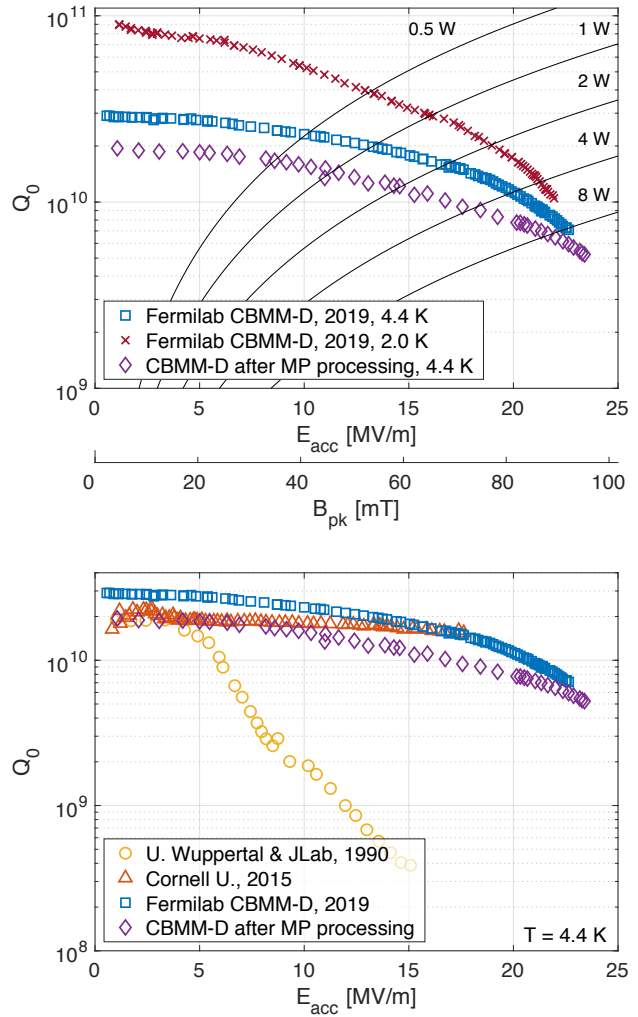

**Fig. 12**: Performance of 1.3 GHz single cell cavity CBMM-D before and after processing (top) and comparison to a selection of previous 1.3-1.5 GHz single cell cavities (data from [23], [32]) at 4.4 K (bottom).

The gradient limit was quench, and there were indications that the quenching might be caused by multipacting. The quench field varied somewhat from cooldown to cooldown but consistently occurred in the multipacting band for a cavity of this type [33]. Temperature mapping and second sound measurements indicated that the quench was highly localized at the equator by polar angle, but widely spread around the cavity azimuthally (see supplemental material for details). To evaluate if it would be possible to process through the possible multipacting, the cavity was reassembled with an antenna with stronger coupling (to better match the cavity after unusually strong $Q_0$-degradation after quench) and retested at VTS. The cavity was quenched repeatedly by applying ~100 W of forward power. This resulted in an increase in the maximum field of the cavity. The cavity was thermal cycled after processing, reaching a maximum gradient of 24 MV/m. The 4.4 K curve is also plotted in the bottom of **Fig. 13** and compared to some previous record Nb$_3$Sn cavity performances (all single cell cavities with frequency close to 1.3 GHz).

The 650 MHz cavity B9AS-AES-002 performed similarly, as shown in **Fig. 13**. It also showed multipacting-like behavior at ~9 MV/m, but without quench. After processing through the multipacting, the cavity achieved a maximum accelerating gradient of 20 MV/m, limited by quench. At the maximum field, the cavity showed a $Q_0$ above $10^{10}$ at 4.4 K. The dissipated power at 10 MV/m is just 1.1 watts at 4.4 K, well within the range of large capacity cryocoolers, demonstrating the potential of coatings with this quality for compact accelerator applications. The performance is substantially improved compared to the limited experience reported in the literature of Nb$_3$Sn cavities with frequency <1 GHz [3].

Following the strong performance of these two cavities, a 9-cell 1.3 GHz cavity, TB9ACC014 was coated. This is the first time a cavity of this type has been coated with Nb$_3$Sn. 9-cell 1.3 GHz cavities are practical accelerator structures ~1 m in length used in state-of-the-art facilities such as the European XFEL and LCLS-II. They are much larger than the single cell cavities typically used in R&D. There is only a small number of past reports of Nb$_3$Sn multi-cell cavities produced via the vapor diffusion coating process, and with limited success in achieving uniform coatings over the length of the structure [4]. For TB9ACC014, it appears that the coating was sufficiently uniform for good performance, as shown in the performance curves at 1.5 K and 4.4 K in **Fig. 14** (note that the plot includes a 0.8 n$\Omega$ correction for losses from each of the stainless steel flanges). While not as strongly performing as the single cell cavities, it shows significant progress towards making Nb$_3$Sn coatings practical for applications. This result is comparable to concurrent efforts by Eremeev et al. on a pair of 5-cell 1.5 GHz cavities, which are smaller in size, but have obtained higher maximum gradients in vertical test [34].

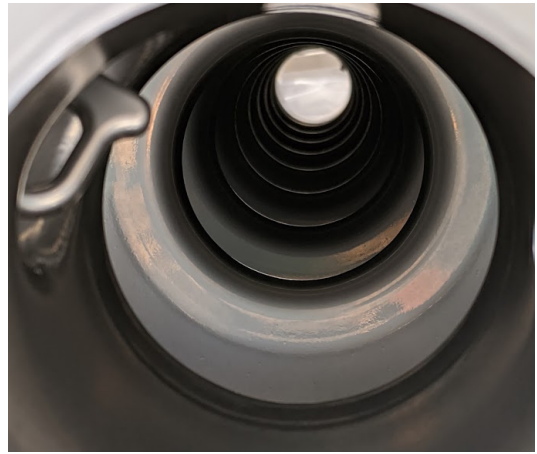

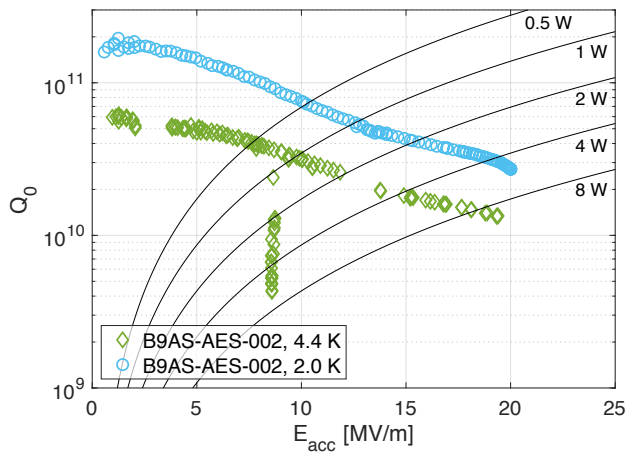

**Fig. 13**: Performance of 650 MHz single cell cavity B9AS-AES-002. The multipacting at 9 MV/m was processed during the test.

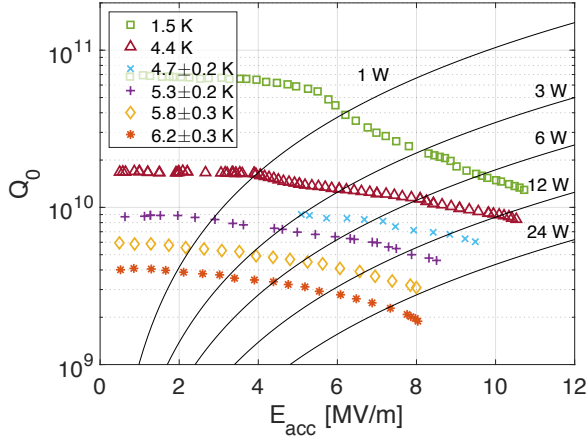

**Fig. 14**: Top: inner surface of the 9-cell 1.3 GHz cavity TB9ACC014. Bottom: performance measured at 1.5 K and 4.4 K in liquid helium and at higher temperatures in cold helium gas (bottom).

Performance curves were also measured at higher temperatures with cooling from cold helium gas. The temperature stability was much less precise than with liquid cooling – **Fig. 14** shows several curves and the corresponding temperature range measured by sensors attached to the outside surface of the cavity. Measuring $Nb_3Sn$ cavities at higher temperatures is useful for developing an optimum operating temperature, which may be significantly higher than for niobium. If cryocoolers are used for cooling, for some types of cavities (e.g. at low frequencies), the optimal temperature may be above 4 K.

TB9ACC014 was electropolished and coated a second time. Its performance after the second coating was very similar to that in the first coating, as shown in **Fig. 15**. This may suggest a defect in the substrate being the performance limit. Considering this possibility, a second 9-cell cavity was coated, TB9AES005. This cavity had even higher maximum field than TB9ACC014, reaching above 15 MV/m before quench, with a $Q_0$ of $9\times10^9$ at 4.4 K. Its performance is also plotted in **Fig. 15** (includes stainless steel flange correction). The bottom of the figure plots the data at 1.5 K (TB9ACC014 test 1) or 2.0 K (TB9ACC014 test 2 and TB9AES005). They are plotted together because typically for our 1.3 GHz $Nb_3Sn$ cavities, BCS resistance at 2.0 K is small compared to residual resistance, so the Q vs E curves at these two temperatures are very similar.

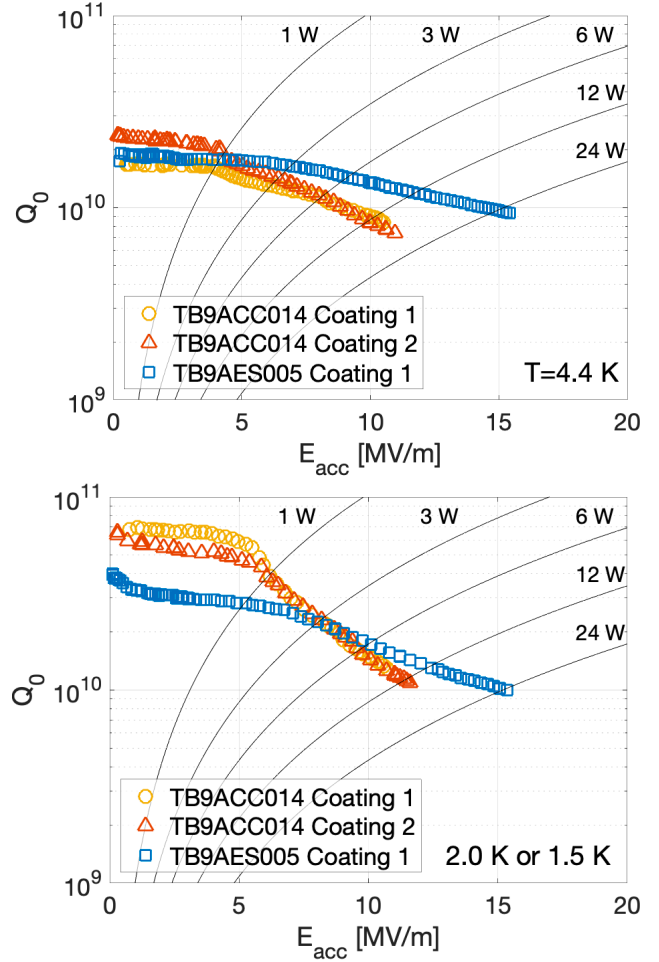

**Fig. 15**: Vertical test performance of $Nb_3Sn$ coated 9-cell cavities TB9ACC014 and TB9AES005. Curves were measured at 4.4 K (top) and 2.0 K or 1.5 K (bottom). Though all three coatings had very similar parameters, TB9AES005 has superior performance, suggesting a possible substrate dependence.

### V. Discussion

The cavity performance presented in the results section is unprecedented for cavities made with SRF materials beyond Nb. This includes two single cell $Nb_3Sn$ cavities with cw accelerating gradients of 20 and 24 MV/m, significantly exceeding the previous highest value of 18 MV/m [1]. It also includes two 9-cell 1.3 GHz $Nb_3Sn$ cavities, which reached 11 and 15 MV/m, an important demonstration on cavities typical of those used in applications. Even close to the maximum gradient, each of these cavities maintained a $Q_0$ on the order of $10^{10}$ at 4.4 K.

Each of these cavities had a shiny surface, not typically seen in $Nb_3Sn$ SRF coatings. It seems likely that this is linked to the strong performance, and additional measurements will be performed in the future to evaluate this possibility. There are reasons to expect that the shiny coatings would improve performance. Microstructural analysis shown in **Fig. 3** and **Fig. 5** reveal that the shiny coatings are correlated with lower surface roughness. Surface roughness is caused by grooves and height differences between grains, **Fig. 6**, and it is suppressed with the smaller grain sizes observed in shiny films. Surface roughness can cause field enhancement and lowering of the energy barrier, causing penetration of magnetic flux at fields lower than the superheating field [20], meaning that a smoother layer may reach higher fields. Simulations in Ref. [35] of type II superconductors using Ginzburg-Landau theory show how grooves in surfaces can be vulnerable to flux penetration. Analyses shown in **Fig. 4** and **Fig. 5** reveal that the shiny coatings are also correlated with a thinner layer. $Nb_3Sn$ has significantly lower thermal conductivity than Nb [21], [36], so a thinner layer may be helpful for preventing thermal runaway, helping to stabilize the temperature on the inner surface, especially at local defects.

$Nb_3Sn$ cavities can be thermally stable even with large amounts of dissipation over the surface, e.g. ~30 W over a single cell cavity [37]. Consider for example, CBMM-D, which the supplemental material shows has local T-map measurements as high as ~0.5 K after quench. However, before quench, the maximum temperature measured by the T-map is only ~0.05 K. If thermal runaway were the key limitation in this test, one might expect that the heating pre-quench would have been higher or that the higher temperature observed by the T-map after quench would have resulted in a significantly lower maximum field. Since neither of these were true, it might indicate that surface roughness is more likely to play an important role than the film thickness, at least in this case.

It is noteworthy that the cavities were all treated with a N-infusion step, but no excess of N was found in the near surface of a witness sample using SIMS in comparison to a non-infused sample. This may suggest that the N-infusion was not the key factor in performance improvement. Continued studies without this step will help to evaluate its importance.

The high temperature nucleation step appears to help with uniformity and density of coverage of islands early in the coating process, based on sample host cavity studies in **Fig. 11**. This is expected to help prevent the growth of pachy regions. On the other hand, it is not yet clear what the key factors are that contributed to the formation of smooth, thin coatings—this is a focus of continued investiations. If these investigations lead to consistent achievement of gradients >20 MV/m in $Nb_3Sn$ cavities, it would be extremely beneficial for accelerator applications.

The films under study in these experiments are bulk films. There have been concerns in the SRF community about flux penetration at fields above the lower critical field into bulk superconductors that, like $Nb_3Sn$, have coherence length significantly smaller than Nb, and therefore may be vulnerable to flux entering at small defects. This has led to efforts to develop multilayer coatings with insulating films between superconducting layers to prevent dendritic flux avalanches [38], [39]. The single cell 1.3 GHz cavity (CBMM-D) reached a peak surface magnetic field of approximately 95 mT, well above the predicted lower critical field of $Nb_3Sn$ [40], [41], strengthening previous assertions that bulk $Nb_3Sn$ cavities can be maintained in the metastable flux-free state [42]. The small coherence length of $Nb_3Sn$ and the presence of small-scale defects in realistic films may eventually make multilayer films advantageous, but the experiments presented here demonstrate that $Nb_3Sn$ films in bulk form can reach fields that are sufficiently large for many SRF applications. Furthermore, if it is determined that film smoothness and/or reduced layer thickness are the key factors in the performance presented in this paper, then continued progress in accelerating gradient in $Nb_3Sn$ bulk films may be enabled by further modification of coating and post-processing steps to improve these factors.

### VI. Conclusions

With the single cell 1.3 GHz cavity, the single cell 650 MHz cavity, and the two 9-cell 1.3 GHz cavities, significant progress is reported on $Nb_3Sn$ SRF cavity development in this paper:

- A record CW accelerating gradient of 24 MV/m for a cavity made with an SRF material other than Nb, significantly higher than the previous highest CW gradient for a $Nb_3Sn$ cavity ~18 MV/m
- Two cavities that exhibit $Q_0 > 1\times10^{10}$ at 20 MV/m and 4.4 K, showing promising performance levels for $Nb_3Sn$ in cryocooler-based compact accelerator applications as well

- as larger-scale accelerator applications where cryogenics are a cost driver;
- $Q_0 \sim 9 \times 10^9$ at 15 MV/m and 4.4 K on a practical cavity structure commonly used in accelerators;
- First demonstration of multipacting being processed in Nb$_3$Sn cavities.

These cavities all had shiny surfaces, atypical for Nb$_3$Sn. Microstructural analysis shows that the shiny films are smoother and thinner than matte films, possibly leading to higher maximum fields due to reduced vulnerability to flux penetration at dips and grooves and/or improved thermal stability. Future work will focus on achieving strong reproducibility, continued progress in maximum field, and further steps towards first applications.

**Acknowledgements**

The authors are grateful for the dedicated efforts of the SRF processing team at FNAL and ANL, the FNAL VTS testing team, and the FNAL machine shop and welding experts. Thanks also for helpful discussions with Anna Grassellino, Sergey Belomestnykh, Hasan Padamsee, Curtis Crawford, Chuck Grimm, Matthias Liepe, Grigory Eremeev, Daniel Hall, Ryan Porter, Uttar Pudasaini, and the FNAL SRF science team. This manuscript has been authored by Fermi Research Alliance, LLC under Contract No. DE-AC02-07CH11359 with the U.S. Department of Energy, Office of Science, Office of High Energy Physics and supported by the primary author's DOE Early Career Award. This work made use of the EPIC, Keck-II, and/or SPID facilities of Northwestern University's NU*ANCE* Center, which received support from the Soft and Hybrid Nanotechnology Experimental (SHyNE) Resource (NSF ECCS-1542205); the MRSEC program (NSF DMR-1121262) at the Materials Research Center; the International Institute for Nanotechnology (IIN); the Keck Foundation; and the State of Illinois, through the IIN.

**References**

[1] S. Posen and D. L. Hall, "Nb3Sn superconducting radiofrequency cavities: fabrication, results, properties, and prospects," *Supercond. Sci. Technol.*, vol. 30, no. 3, p. 033004, (2017).

[2] A. Grassellino, A. Romanenko, D. Bice, O. Melnychuk, A. C. Crawford, S. Chandrasekaran, Z. Sung, D. A. Sergatskov, M. Checchin, S. Posen, M. Martinello, and G. Wu, "Accelerating fields up to 49 MV/m in TESLA-shape superconducting RF niobium cavities via 75C vacuum bake," pp. 2–6, (2018).

[3] G. Arnolds-Mayer and E. Chiaveri, "On a 500 MHz single cell cavity with Nb3Sn surface," in *Proceedings of The Third Workshop on RF Superconductivity*, (1986).

[4] M. Peiniger, M. Hein, N. Klein, G. Müller, H. Piel, and P. Thuns, "Work on Nb3Sn cavities at Wuppertal," in *Proceedings of The Third Workshop on RF Superconductivity*, (1988).

[5] M. K. Transtrum, G. Catelani, and J. P. Sethna, "Superheating field of superconductors within Ginzburg-Landau theory," *Phys. Rev. B*, vol. 83, no. 9, p. 094505, (2011).

[6] W. J. Schneider, P. Kneisel, and C. H. Rode, "Gradient optimization for SC CW accelerators," *Proc. Part. Accel. Conf. 2003*, pp. 2863–2865, (2003).

[7] R. Kephart, B. Chase, I. Gonin, A. Grassellino, S. Kazakov, T. Khabiboulline, S. Nagaitsev, R. Pasquinelli, S. Posen, O. Pronitchev, A. Romanenko, and V. Yakovlev, "SRF, Compact Accelerators for Industry & Society," *Proc. SRF2015*, p. FRBA03, (2015).

[8] R. C. Dhuley, R. Kostin, O. Prokofiev, M. I. Geelhoed, T. H. Nicol, S. Posen, J. C. T. Thangaraj, T. K. Kroc, and R. D. Kephart, "Thermal Link Design for Conduction Cooling of SRF Cavities Using Cryocoolers," *IEEE Trans. Appl. Supercond.*, vol. 29, no. 5, pp. 1–5, (2019).

[9] G. Ciovati, J. Anderson, B. Coriton, J. Guo, F. Hannon, L. Holland, M. Lesher, F. Marhauser, J. Rathke, R. Rimmer, T. Schultheiss, and V. Vylet, "Design of a cw, low-energy, high-power superconducting linac for environmental applications," *Phys. Rev. Accel. Beams*, vol. 21, no. 9, p. 91601, (2018).

[10] A. Godeke, "A review of the properties of Nb3Sn and their variation with A15 composition, morphology and strain state," *Supercond. Sci. Technol.*, vol. 19, no. 8, pp. R68–R80, (2006).

[11] E. Saur and J. Wurm, "Preparation und Supraleitungseigenschaften von Niobdrahtproben mit Nb3Sn-Uberzug," *Naturwissenschaften*, vol. 49, no. 6, pp. 127–128, (1962).

[12] B. Hillenbrand, "Superconducting Nb3Sn

cavities with high quality factors and high critical flux densities," *J. Appl. Phys.*, vol. 47, no. 9, p. 4151, (1976).

[13] D. A. Rudman, F. Hellman, R. H. Hammond, and M. R. Beasley, "A15 Nb-Sn tunnel junction fabrication and properties," *J. Appl. Phys.*, vol. 55, no. 10, pp. 3544–3553, (1984).

[14] G. Eremeev, C. E. Reece, M. J. Kelley, U. Pudasaini, and J. R. Tuggle, "Progress with multi-cell Nb3Sn cavity development linked with sample matierals characterization," in *Proceedings of the Seventeenth International Conference on RF Superconductivity, Whistler, Canada*, (2015), p. TUBA05.

[15] Z. Yang, Y. He, H. Guo, C. Li, P. Xiong, M. Lu, S. Zhang, Z. Lin, T. Tan, and S. Zhang, "Development of Nb3Sn cavity coating at IMP," *Proc. Ninteenth Int. Work. RF Supercond. Dresden, Ger.*, (2019).

[16] Y. Trenikhina, S. Posen, A. Romanenko, M. Sardela, J. Zuo, D. L. Hall, and M. Liepe, "Performance-defining properties of Nb 3 Sn coating in SRF cavities," *Supercond. Sci. Technol.*, vol. 31, no. 1, p. 015004, (2018).

[17] J. Lee, S. Posen, Z. Mao, Y. Trenikhina, K. He, D. L. Hall, M. Liepe, and D. N. Seidman, "Atomic-scale analyses of Nb3Sn on Nb prepared by vapor diffusion for superconducting radiofrequency cavity applications: A correlative study," *Supercond. Sci. Technol.*, vol. 32, no. 2, (2019).

[18] T. Spina, B. M. Tennis, J. Lee, D. N. Seidman, and S. Posen, "Development and Understanding of Nb3Sn films for radiofrequency applications through a sample-host 9-cell cavity," *arxiv: 2006.13407*, (2020).

[19] J. Lee, Z. Mao, K. He, Z. H. Sung, T. Spina, S. Il Baik, D. L. Hall, M. Liepe, D. N. Seidman, and S. Posen, "Grain-boundary structure and segregation in Nb3Sn coatings on Nb for high-performance superconducting radiofrequency cavity applications," *Acta Mater.*, vol. 188, pp. 155–165, (2020).

[20] R. Porter, D. L. Hall, M. Liepe, and J. Maniscalco, "Surface Roughness Effect on the Performance of Nb3Sn Cavities," *28th Linear Accel. Conf.(LINAC'16), East Lansing, MI, USA, 25-30 Sept. 2016*, pp. 129–132, (2017).

[21] G. D. Cody and R. W. Cohen, "Thermal Conductivity of Nb3Sn," *Rev. Mod. Phys.*, vol. 36, no. 1, pp. 121–123, (1964).

[22] B. Hillenbrand, H. Martens, H. Pfister, K. Schnitzke, and Y. Uzel, "Superconducting Nb3Sn cavities with high microwave qualities," *IEEE Trans. Magn.*, vol. 13, no. 1, pp. 491–495, (1977).

[23] D. L. Hall, "New insights into the limitations on the efficiency and achievable gradients in Nb3Sn SRF cavities," Cornell University, (2017).

[24] S. . Mucklejohn and N. . O'Brien, "The vapour pressure of tin(II) chloride and the standard molar Gibbs free energy change for formation of SnCl2(g) from Sn(g) and Cl2(g)," *J. Chem. Thermodyn.*, vol. 19, no. 10, pp. 1079–1085, (1987).

[25] B. Hillenbrand, "The Preparation of Superconducting Nb3Sn Surfaces for RF applications," in *Proceedings of the First Workshop on RF Superconductivity*, (1980).

[26] A. Grassellino, A. Romanenko, Y. Trenikhina, M. Checchin, M. Martinello, O. S. Melnychuk, S. Chandrasekaran, D. A. Sergatskov, S. Posen, A. C. Crawford, S. Aderhold, and D. Bice, "Unprecedented quality factors at accelerating gradients up to 45 MVm-1in niobium superconducting resonators via low temperature nitrogen infusion," *Supercond. Sci. Technol.*, vol. 30, no. 094004, (2017).

[27] U. Pudasaini, G. V. Eremeev, C. E. Reece, J. Tuggle, and M. J. Kelley, "Effect of Deposition Temperature and Duration on Nb3Sn Diffusion Coating," *Ninth Int. Part. Accel. Conf. Vancouver, BC, Canada*, p. THPAL130, (2018).

[28] R. Ridgway and S. Posen, "Simulating Nb3Sn Coating Process Inside SRF Cavities," *Fermilab Summer Internsh. Sci. Technol. Rep.*, pp. 1–4, (2017).

[29] M. D. Abràmoff, P. J. Magalhães, and S. J. Ram, "Image processing with ImageJ Part II," *Biophotonics Int.*, vol. 11, no. 7, pp. 36–43, (2005).

[30] U. Pudasaini, G. V. Eremeev, C. E. Reece, J. Tuggle, and M. J. Kelley, "Initial growth of tin on niobium for vapor diffusion coating of Nb 3 Sn," *Supercond. Sci. Technol.*, vol. 32, no. 4, (2019).

[31] H. Padamsee, J. Knobloch, and T. Hays, *RF superconductivity for accelerators*. New York: Wiley-VCH, (2008).

[32] G. Müller, H. Piel, J. Pouryamout, P. Boccard, and P. Kneisel, "Status and Prospects of Nb3Sn Cavities for Superconducting Linacs," in *Proceedings of the Workshop on Thin Film Coating Methods for Superconducting*

*Accelerating Cavities*, (2000).

[33] P. Ylä-Oijala, "Electron Multipacting in TESLA Cavities and Input Couplers," *Part. Accel.*, vol. 63, pp. 105–137, (1999).

[34] G. Eremeev, W. Clemens, K. Macha, C. E. Reece, A. M. Valente-Feliciano, S. Williams, U. Pudasaini, and M. Kelley, "Nb3Sn multicell cavity coating system at JLAB," pp. 1–6, (2020).

[35] A. R. Pack, J. Carlson, S. Wadsworth, and M. K. Transtrum, "Vortex nucleation in superconductors within time-dependent Ginzburg-Landau theory in two and three dimensions: Role of surface defects and material inhomogeneities," *Phys. Rev. B*, vol. 101, no. 14, pp. 1–10, (2020).

[36] F. Koechlin and B. Bonin, "Parametrization of the niobium thermal conductivity in the superconducting state," *Supercond. Sci. Technol.*, vol. 9, no. 6, pp. 453–460, (1996).

[37] S. Posen, "Understanding and Overcoming Limitation Mechanisms in Nb3Sn Superconducting RF Cavities," Cornell University, (2015).

[38] A. Gurevich, "Enhancement of rf breakdown field of superconductors by multilayer coating," *Appl. Phys. Lett.*, vol. 88, no. 1, p. 012511, (2006).

[39] A. Gurevich, "Maximum screening fields of superconducting multilayer structures," *AIP Adv.*, vol. 5, no. 1, p. 017112, (2015).

[40] T. P. Orlando, E. J. McNiff, S. Foner, and M. R. Beasley, "Critical fields, Pauli paramagnetic limiting, and material parameters of Nb3Sn and V3Si," *Phys. Rev. B*, vol. 19, no. 9, pp. 4545–4561, (1979).

[41] A. Godeke, "Performance boundaries in Nb3Sn superconductors," University of Twente, Enschede, The Netherlands, (2005).

[42] S. Posen, M. Liepe, and D. L. Hall, "Proof-of-principle demonstration of Nb3Sn superconducting radiofrequency cavities for high Q0 applications," *Appl. Phys. Lett.*, vol. 106, no. 8, p. 082601, (2015).

## Supplemental material: Measurements suggesting that quench in single cell 1.3 GHz cavity CBMM-D was induced by multipacting

The single cell 1.3 GHz cavity CBMM-D was vertically tested with a temperature mapping system (T-map), an array of hundreds of resistance temperature detectors over the surface of the cavity. Just before quench, the T-map showed a wide distribution of heating over the surface of the cavity, as shown in Fig. A1.

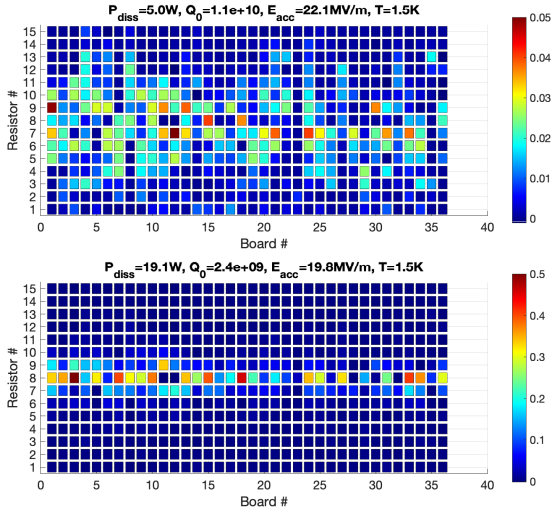

Fig A1: Temperature maps recorded at ~20 MV/m before quench had occurred (top) and after quench had occurred (bottom). Note the order of magnitude difference in color scales. Colorbar units are K.

Fig. A1 also shows a T-map measured after quench, showing that there was fairly high heating, with maximum readings ~0.5 K with the cavity in steady state at 20 MV/m. This amount of heating is not unusual for a $Nb_3Sn$ cavity after quench, and is usually interpreted as being caused by flux trapped at the quench location, with the flux being generated by thermocurrents from large thermal gradients around the quench site while it is briefly normal conducting before cooling down again. On the other hand, the heating pattern is atypical. There was strong heating very highly localized at the equator of the cavity (the resistors each cover an area of approximately 1 cm$^2$, and the heating was localized within a single resistor by polar angle), but widely distributed over the entire $2\pi$ of azimuth around the equator. Cavity degradation phenomena such as local defects and field emission are expected to be localized or have heating patterns along azimuthal, rather than polar lines [1]. For this shape of cavity, the maximum peak surface magnetic field is widely distributed over a large surface area (covering approximately resistor numbers 4-12 on each board of the T-map), and is actually slightly lower at the equator, so this heating distribution is unlikely to be due to this part of the cavity reaching a specific peak magnetic field. However, two-point multipacting in the multipacting band of the elliptical cell is expected to be highly localized at the equator.

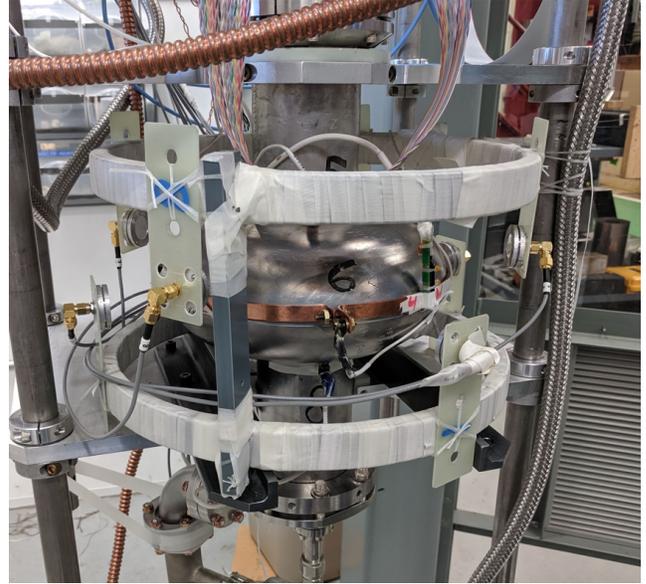

Fig. A2: Instrumentation around cavity for vertical test with second sound transducers. The transducers are the cylindrical metallic devices attached to G-10 boards, which themselves are attached to coils around the cavity that are used to cancel the longitudinal ambient magnetic field.

To further investigate whether quench might be caused by multipacting, the cavity was vertically tested again, this time with the T-map removed and second-sound transducers added. When quench occurs in superfluid helium, a temperature-entropy wave travels to these second-sound transducers at a velocity of ~20 m/s, making it possible to triangulate back to the quench location by correlating signals from several sensors [2]. For this experiment, eight second sound transducers were placed near the equator of the cavity at different azimuthal angles, as shown in Fig. A2. When quench occurred, it was captured by all eight sensors. If it had occurred in one location (e.g. a defect), the signal would have appeared on the closest transducer soonest, then appeared on more distant sensors in turn. Instead, the signal appeared on all the sensors at nearly the same time. The small differences in time were compared to measurements of the distances of the transducers from the equator. The comparison, in Fig. A3, is consistent with the entire equator becoming normal conducting within a very short period of time, such that the time it takes for the signal to appear on a given transducer is approximately equal to the distance between it and the equator divided by the speed of second sound propagation.

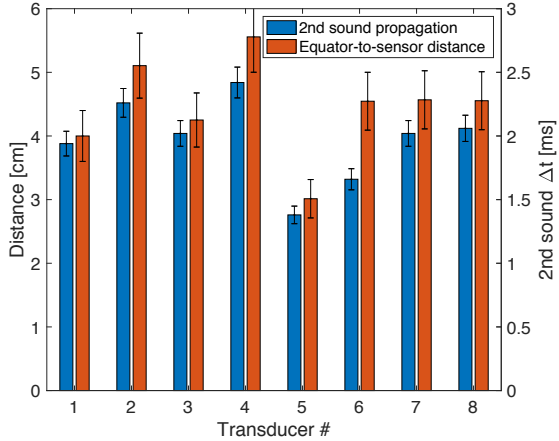

Fig. A3: Agreement between the distance betweeen each transducer and the equator and the time taken for the second sound signal to propagate. This suggests that practically the entire equator rapidly and briefly became normal conducting during the quench.

The second sound measurement is consistent with that of the T-map, showing that the quench is widely distributed around the equator. However the mechanism behind such a quench it isn't immediately obvious. A possible explanation is that the trigger for multipacting was less probably than usual, such that the cavity was able to operate for some time in the multipacting band, in conditions that would be favorable for two-point multipacting at the equator, but without a trigger for multipacting occurring. Once the multipacting was finally triggered, it happened intensely. There would be no reason for it to be localized in azimuthal angle, but rather would spread around the equator, leading to large amounts of dissipation from electron impacts and eventually quench. Because of the timescales involved, there would be time for many RF cycles with multipacting electrons before the RF field was reduced to zero. This scenario would be consistent with the observations but would also be atypical. The behavior may also be caused by a different mechanism, but it would have to explain both the unusual heating pattern and the lack of pre-heating at the equator before quench occurred (i.e. as Fig. A1 shows, the heating was approximately equivalent at the equator as in other high magnetic field areas before quench).

In both of these tests, the cavity could only be quenched 1-2 times before the $Q_0$ had degraded so much that it was well below the $Q_{ext}$ of the input coupler, and there was not sufficient RF power available from the amplifier to bring the cavity to quench. It was therefore reassembled with an antenna with stronger coupling, as described in the main text. Radiation was observed during quenching, and the cavity was processed to a higher maximum field, both of which are also consistent with multipacting.

After the test described in the main body of the paper in which the multipacting in the cavity was processed, it was again put into the dewar, this time with a T-map. T-maps were measured before and after quench occurred. After the first quench, very little $Q_0$ degradation occurred. The T-map showed heating localized in one area, suggesting a non-multipactinig-related quench. The cavity was then quenched a few times, and the T-map showed wide heating around the equator, suggesting multipacting related quench, shown in Fig. A4. The heating from flux trapped during the first quench was still present. The cavity was then held at a field in the low 20 MV/m range for several minutes, and quenches sporatically occurred, again consistent with multipacting.

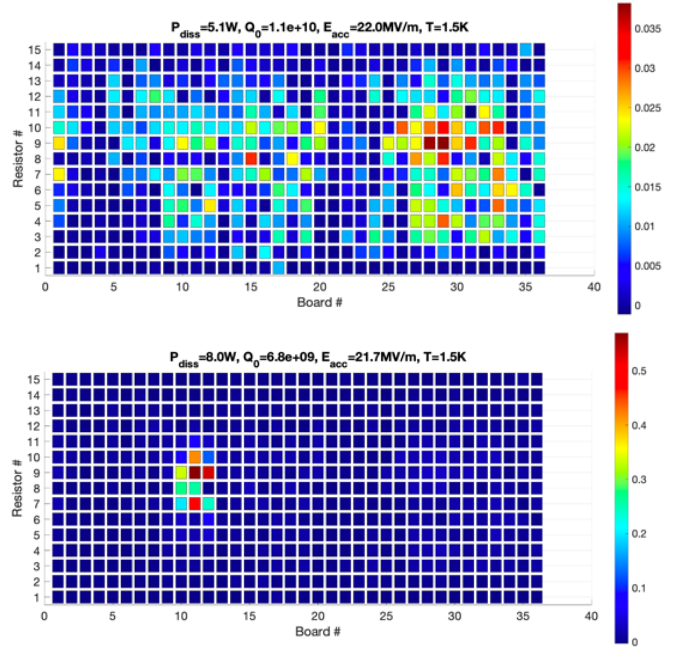

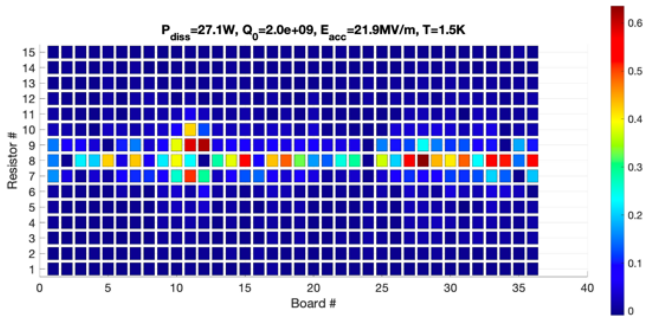

Fig. A4: Temperature maps recorded before quench had occurred (top), after expected non-multipacting-related quench (middle), and after expected multipacting-related quench (bottom). Note the order of magnitude difference in color scales. Colorbar units are K.

**References**

[1]  J. Knobloch, "Advanced thermometry studies of superconducting RF cavities," Ph.D. Thesis, Cornell University, (1997).

[2]  Z. A. Conway, D. L. Hartill, H. S. Padamsee, and E. N. Smith, "Oscillating superleak transducers for quench detection in superconducting ILC cavities cooled with HE-II," *Proc. 24th Linear Accel. Conf. LINAC 2008*, pp. 863–865, (2009).